# Supporting SPARQL Update Queries in RDF-XML Integration [*]

Nikos Bikakis[1][†]  Chrisa Tsinaraki[2]  Ioannis Stavrakantonakis[3]
Stavros Christodoulakis[4]

[1] NTU Athens & R.C. ATHENA, Greece
[2] EU Joint Research Center, Italy
[3] STI, University of Innsbruck, Austria
[4] Technical University of Crete, Greece

**Abstract.** The *Web of Data* encourages organizations and companies to publish their data according to the Linked Data practices and offer SPARQL endpoints. On the other hand, the dominant standard for information exchange is XML. The *SPARQL2XQuery Framework* focuses on the automatic translation of SPARQL queries in XQuery expressions in order to access XML data across the Web. In this paper, we outline our ongoing work on supporting update queries in the RDF–XML integration scenario.

## 1 Introduction

The *SPARQL2XQuery Framework*, that we have previously developed [6], aims to bridge the heterogeneity issues that arise in the consumption of XML-based sources within Semantic Web. In our working scenario, mappings between RDF/S–OWL and XML sources are automatically derived or manually specified. Using these mappings, the SPARQL queries are translated on the fly into XQuery expressions, which access the XML data. Therefore, the current version of SPARQL2XQuery provides read-only access to XML data. In this paper, we outline our ongoing work on extending the SPARQL2XQuery Framework towards supporting SPARQL update queries.

Both SPARQL and XQuery have recently standardized their update operation semantics in the SPARQL 1.1 and XQuery Update Facility, respectively. We have studied the correspondences between the update operations of these query languages, and we describe the extension of our mapping model and the SPARQL-to-XQuery translation algorithm towards supporting SPARQL update queries.

Similarly to the motivation of our work, in the RDB–RDF interoperability scenario, D2R/Update [1] (a D2R extension) and OntoAccess [2] enable SPARQL update queries over relational databases. Regarding the XML–RDB–RDF interoperability scenario [5], the work presented in [3] extends the XSPARQL language [4] in order to support update queries.

---
[*] This paper appears in 13th International Semantic Web Conference (ISWC '14).
[†] This work is partially supported by the EU/Greece funded KRIPIS: MEDA Project

## 2 Translating SPARQL Update Queries to XQuery

This section describes the translation of SPARQL update operations into XQuery expressions using the XQuery Update Facility. We present how similar methods and algorithms previously developed in the SPARQL2XQuery Framework can be adopted for the update operation translation. For instance, graph pattern and triple pattern translation are also used in the update operation translation. Note that, due to space limitations, some issues are presented in a simplified way in the rest of this section and several details are omitted.

Table 1 presents the SPARQL update operations and summarizes their translation in XQuery. In particular, there are three main categories of SPARQL update operations a) Delete Data; b) Insert Data; and c) Delete/Insert. For each update operation, a simplified SPARQL syntax template is presented, as well as the corresponding XQuery expressions. In SPARQL context, we assume the following sets, let *tr* be an *RDF triple* set, *tp* a *triple pattern* set, *trp* a set of triples and/or triple patterns, and *gp* a *graph pattern*. Additionally, in XQuery, we denote as $xE_W$, $xE_I$ and $xE_D$ the sets of *XQuery expressions* (i.e., FLOWR expressions) that have resulted from the translation of the graph pattern included in the Where, Insert and Delete SPARQL clauses, respectively. Let $xE$ be a set of XQuery expressions, $xE(\$v_1, \$v_2,\ldots \$v_n)$ denote that $xE$ are using (as input) the values assigned to XQuery variables $\$v_1, \$v_2,\ldots \$v_n$. Finally, *xn* denotes an *XML fragment*, i.e., a set of XML nodes, and *xp* denotes an *XPath expression*.

**Table 1.** Translation of the SPARQL Update Operations in XQuery

| SPARQL Update Operation | Syntax Template [1] | Translated XQuery Expressions |
|---|---|---|
| DELETE DATA | Delete data{<br>  *tr*<br>} | **delete nodes collection**("http://dataset...")/$xp_1$<br>...<br>**delete nodes collection**("http://dataset...")/$xp_n$ |
| INSERT DATA | Insert data{<br>  *tr*<br>} | **let** $\$n_1 := xn_1$<br>...<br>**let** $\$n_n := xn_n$<br>**let** $\$data_1 := (\$n_k, \$n_m,\ldots)$  // $k, m,\ldots \in [1,n]$<br>...<br>**let** $\$data_p := (\$n_j, \$n_v,\ldots)$  // $j, y,\ldots \in [1,n]$<br>**let** $\$insert\_location_1 := \textbf{collection}$("http://xmldataset...")/$xp_1$<br>...<br>**let** $\$insert\_location_p := \textbf{collection}$("http://xmldataset...")/$xp_p$<br>**return**(<br>  **insert nodes** $\$data_1$ **into** $\$insert\_location_1$ ,<br>  ...<br>  **insert nodes** $\$data_p$ **into** $\$insert\_location_p$<br>) |
| DELETE / INSERT | (a) **Delete**{<br>  *trp*<br>}**Where**{<br>  *gp*<br>}<br><br>(b) **Insert**{<br>  *trp*<br>}**Where**{<br>  *gp*<br>}<br><br>(c) **Delete**{<br>  *trp*<br>}**Insert**{<br>  *trp*<br>}**Where**{<br>  *gp*<br>} | (a)<br>**let** $\$where\_gp := xE_W$<br>**let** $\$delete\_gp := xE_D(\$where\_gp)$<br>**return delete nodes** $\$delete\_gp$<br><br>(c)<br>Translate Delete Where same as (a), then translate Insert Where same as (b) | (b) **let** $\$where\_gp := xE_W$<br>**let** $\$insert\_location_1 := xp_1$<br>**for** $\$it_1$ **in** $\$insert\_location_1$<br>$xE_I(\$where\_gp, \$it_1)$<br>**return insert nodes into** $\$it_1$<br>...<br>**let** $\$where\_gp := xE_W$<br>**let** $\$insert\_location_n := xp_n$<br>**for** $\$it_n$ **in** $\$insert\_location_n$<br>$xE_I(\$where\_gp, \$it_n)$<br>**return insert nodes into** $\$it_n$ |

[1] For simplicity, the WITH, GRAPH and USING clauses are omitted.

In the following examples, we assume that an RDF source has been mapped to an XML source. In particular, we assume the example presented in [6], where an RDF and an XML source describing persons and students have been mapped. Here, due to space limitation, we just outline the RDF and XML concepts, as well as the mappings that are involved in the following examples. In RDF, we have a class Student having several datatype properties, i.e., FName, E-mail, Department, GivenName, etc. In XML, we have an XML complex type Student_type, having an attribute SSN and several simple elements, i.e., FirstName, Email, Dept, GivenName etc. Based on the XML structure, the students' elements appear in the \Persons\Student path. We assume that the Student class has been mapped to the Student_type and the RDF datatype properties to the similar XML elements.

**Delete Data.** The Delete Data SPARQL operation removes a set of triples from RDF graphs. This SPARQL operation can be translated in XQuery using the Delete Nodes XQuery operation. Specifically, using the predefined mappings, the set of triples *tr* defined in the SPARQL Delete Data clause is transformed (using a similar approach such as the *BGP2XQuery* algorithm [6]) in a set of XPath expressions XP. For each $xp_i \in$ XP an XQuery Delete Nodes operation is defined.

**Example 1.** In this example, two RDF triples are deleted from an RDF graph. In addition to the mappings described above, we assume that the person "*http://rdf.gr/person1209*" in RDF data has been mapped to the person "*/Persons/Student[.@SSN=1209]*" in XML data.

*SPARQL Delete Data query* ⇒ *Translated XQuery query*

**Delete data**{
   <http://rdf.gr/person1209> ns: FName "John" .
   <http://rdf.gr/person1209> ns:E-mail "john@smith.com".
}

**delete nodes collection**("http://xml.gr")/Persons/Student[.@SSN=1209]/FirstName[.= "John"]
**delete nodes collection**("http://xml.gr")/Persons/Student[.@SSN=1209]/Email[.= "John@smith.com"]

**Insert Data.** The Insert Data SPARQL operation, adds a set of new triples in RDF graphs. This SPARQL operation can be translated in XQuery using the Insert Nodes XQuery operation. In the Insert Data translation, the set of triples *tr* defined in SPARQL are transformed into XML node sets $xn_i$, using the predefined mappings. In particular, a set of Let XQuery clauses is used to build the XML nodes and define the appropriate node nesting and grouping. Then, the location of the XML node insertion can be easily determined considering the triples and the mappings. Finally, the constructed nodes are inserted in their location of insertion using the XQuery Insert nodes clause.

**Example 2.** In this example, the RDF triples deleted in the previous example are re-inserted in the RDF graph.

*SPARQL Insert Data query* ⇒ *Translated XQuery query*

**Insert data**{
   <http://rdf.gr/person1209> ns:FName "John" .
   <http://rdf.gr/person1209> ns:E-mail "john@smith.com".
}

**let** $n1 := <FirstName>John</FirstName>
**let** $n2 := <Email>john@smith.com</Email>
**let** $data1 := ($n1, $n2)
**let** $insert_location1 := **collection**("http://xml.gr")/Persons/Student[.@SSN=1209]
**return insert nodes** $data$_1$ **into** $insert_location1

**Insert / Delete.** The Delete/Insert SPARQL operations are used to remove and/or add a set of triples from/to RDF graphs, using the bindings that resulted from the evaluation of the graph pattern defined in the Where clause. According to the SPARQL 1.1 semantics, the Where clause is the first one that is evaluated. Then, the Delete/Insert clause is applied over the produced results. Especially, in case, that both Delete and Insert operations exist, the deletion is performed before the insertion, and the Where clause is evaluated once. The Delete and the Insert SPARQL operations can be translated to XQuery using the Delete Nodes and Insert Nodes operations, respectively. In brief, initially the graph pattern used in the Where clause is translated to XQuery expressions $xE_W$ (similarly as in the *GP2XQuery* algorithm [6]). Then, the graph pattern used in the Delete/Insert clause is translated to XQuery expressions $xE_D/xE_I$ (as it is also in the *BGP2XQuery* algorithm [6]) using also the bindings that resulted from the evaluation of $xE_W$.

**Example 3.** In this example, the Where clause selects all the students studying in a computer science (CS) department. Then, the Delete clause deletes all the triples that match with its triple patterns, using the *?student* bindings determined from the Where clause. In particular, from all the retrieved students (i.e., CS students), the students which have as first name the name "*John*" should be deleted.

*SPARQL Delete query* $\Rightarrow$ *Translated XQuery query*
**Delete**{
  ?student ns:FName "John" .
}**Where**{
  ?student ns:Department "CS" .
}

**let** $where_gp := **collection**("http://xml.gr")/Persons/Student[./Dept="CS"]
**let** $delete_gp := $where_gp[./FirstName="John"]
**return delete nodes** $delete_gp

**Example 4.** In this example, the Where clause selects all the students studying in a CS department, as well as their first names. Then, the Insert clause creates new triples according to its triple patterns, using the *?student* and *?name* bindings determined from the Where clause. In particular, a new triple having as predicate *"ns:GivenName"* and as object the first name of the *?student*, is inserted for each *?student*.

*SPARQL Insert query* $\Rightarrow$ *Translated XQuery query*
**Insert**{
  ?student ns:GivenName ?name .
}**Where**{
  ?student ns:FName ?name .
  ?student ns:Department "CS" .
}

**let** $where_gp := **collection**("http://xml.gr")/Persons/Student[./Dept="CS"]
**let** $insert_location1 := $where_gp
**for** $it1 **in** $insert_location1
**let** $insert_gp1 := <GivenName>{fn:string($it1/FirstName)}</GivenName>
**return insert nodes** $insert_gp1 **into** $it1

## References

1. Eisenberg V., Kanza Y.: "D2RQ/update: updating relational data via virtual RDF". In WWW 2012
2. Hert M., Reif G., Gall H. C.: "Updating relational data via SPARQL/update". In EDBT/ICDT Workshops 2010.
3. Ali M.I., Lopes N., Friel O., Mileo A.: "Update Semantics for Interoperability among XML, RDF and RDB". In APWeb 2013
4. Bischof S., Decker S., Krennwallner T., Lopes N., Polleres A.: "Mapping between RDF and XML with XSPARQL". J. Data Semantics 1(3), (2012)
5. Bikakis N., Tsinaraki C., Gioldasis N., Stavrakantonakis I., Christodoulakis S.: "The XML and Semantic Web Worlds: Technologies, Interoperability and Integration. A survey of the State of the Art". In Semantic Hyper/Multi-media Adaptation: Schemes and Applications, Springer 2013
6. Bikakis N., Tsinaraki C., Stavrakantonakis I., Gioldasis N., Christodoulakis S.: "The SPARQL2XQuery Interoperability Framework". World Wide Web Journal (WWWJ), 2014